\documentclass{llncs} %splncs03
\usepackage{epsf}
\usepackage[colorinlistoftodos]{todonotes}
\usepackage{siunitx}
\usepackage{verbatim}
\usepackage{cite}
\usepackage{subcaption}
\usepackage[utf8]{inputenc}
\usepackage[T1]{fontenc}
\usepackage{fourier, erewhon}
\usepackage{geometry}
\usepackage{soul}
\usepackage{array, floatrow, tabularx, makecell, booktabs}
\usepackage{lineno}
\setcellgapes{3pt}

\newcommand{\nelli}[1]{\textsc{Nelli}}

\usepackage{hyperref}
\newcommand{\footurl}[1]{\footnote{\small\url{#1}}}

\title{Communication Now and Then: Analyzing the Republic of Letters as a Communication Network}
\author{
Javier Ureña-Carrion \inst{1}\orcidID{0000-0003-2763-9747} \and 
Petri Leskinen\inst{2}\orcidID{0000-0003-2327-6942}  \and 
Jouni Tuominen\inst{2,3}\orcidID{0000-0003-4789-5676} \and Charles van den Heuvel\inst{4} \orcidID{0000-0001-9638-400X} \and 
Eero Hyv\"{o}nen\inst{2,3}\orcidID{0000-0003-1695-5840} \and 
Mikko Kivelä\inst{1}\orcidID{0000-0003-2049-1954} 
}

\institute{Complex Systems Group, Aalto University, Finland \\
\and 
Semantic Computing Research Group (SeCo), Aalto University, Finland\\
https://seco.cs.aalto.fi
\and
HELDIG -- Helsinki Centre for Digital Humanities, University of Helsinki, Finland\\
https://heldig.fi\\
\and
Huygens Institute for the History of the Netherlands: Amsterdam, NL
}

\begin{document}

\maketitle
\begin{abstract}
Huge advances in understanding patterns of human communication, and the underlying social networks where it takes place, have been made recently using massive automatically recorded data sets from digital communication, such as emails and phone calls. However, it is not clear to what extent these results on human behaviour are artefacts of contemporary communication technology and culture and if the fundamental patterns in communication have changed over history.
This paper presents an analysis of historical epistolary metadata with the aim of comparing the underlying historical communication patterns with those of contemporary communication. 
Our work uses a new epistolary dataset containing metadata on over \num{150 000} letters sent between the 16th and 19th centuries.
The analyses indicate striking resemblances between contemporary and epistolary communication network patterns, including dyadic interactions and ego-level behaviour. Despite these positive findings, certain aspects of the letter datasets are insufficient to corroborate other similarities or differences for these communication networks. 
\end{abstract}

\section{Introduction}
New communication technologies and their digital traces have revolutionized the social sciences by providing new tools for understanding human behaviour \cite{lazer2020computational,lazer2021meaningful}, and permitting to model society as a network of diverse and multifaceted relationships~\cite{Onnela2007, Candia2008, Borgatti2009, JariTemp, Miritello2013b}.
This has resulted in important insights regarding patterns of human communication~\cite{Miritello2012a, MiritelloTimeAllocation} and the structure of social networks ~\cite{Barabsi2002b, Onnela2007}. %, as well as related social phenomena such as disease spreading~\cite{Kivela2012}, perception of social minorities~\cite{Karimi2018} and polarization~\cite{chen2020polarization}.
However, these analyses have been mainly described with data from contemporary human societies such as mobile phone calls, emails and online platforms.
%These analyses have been mainly described with data from contemporary human societies---mobile phone calls, emails and online platforms---, and far fewer studies have focused on historical communication mediums \cite{Oliveira2005, Malmgren2009}.
Although these phenomena have been observed in a myriad of contemporary communication media, 
the extent to which these phenomena are a product of current technologies and practices, or whether they are also present in other historical contexts and thus a characteristic of human behaviour is unclear as this question has mostly been analysed from theoretical perspectives~\cite{Innis1951, McLuhan1988}. In this paper, we study a large historical epistolary dataset under a behavioural framework, and investigate whether letters ---a major communication channel during the previous centuries--- reflect patterns of human behaviour known to be present in contemporary communication. 

\begin{figure}[ht]
    \centering
    \includegraphics[width=.8\textwidth]{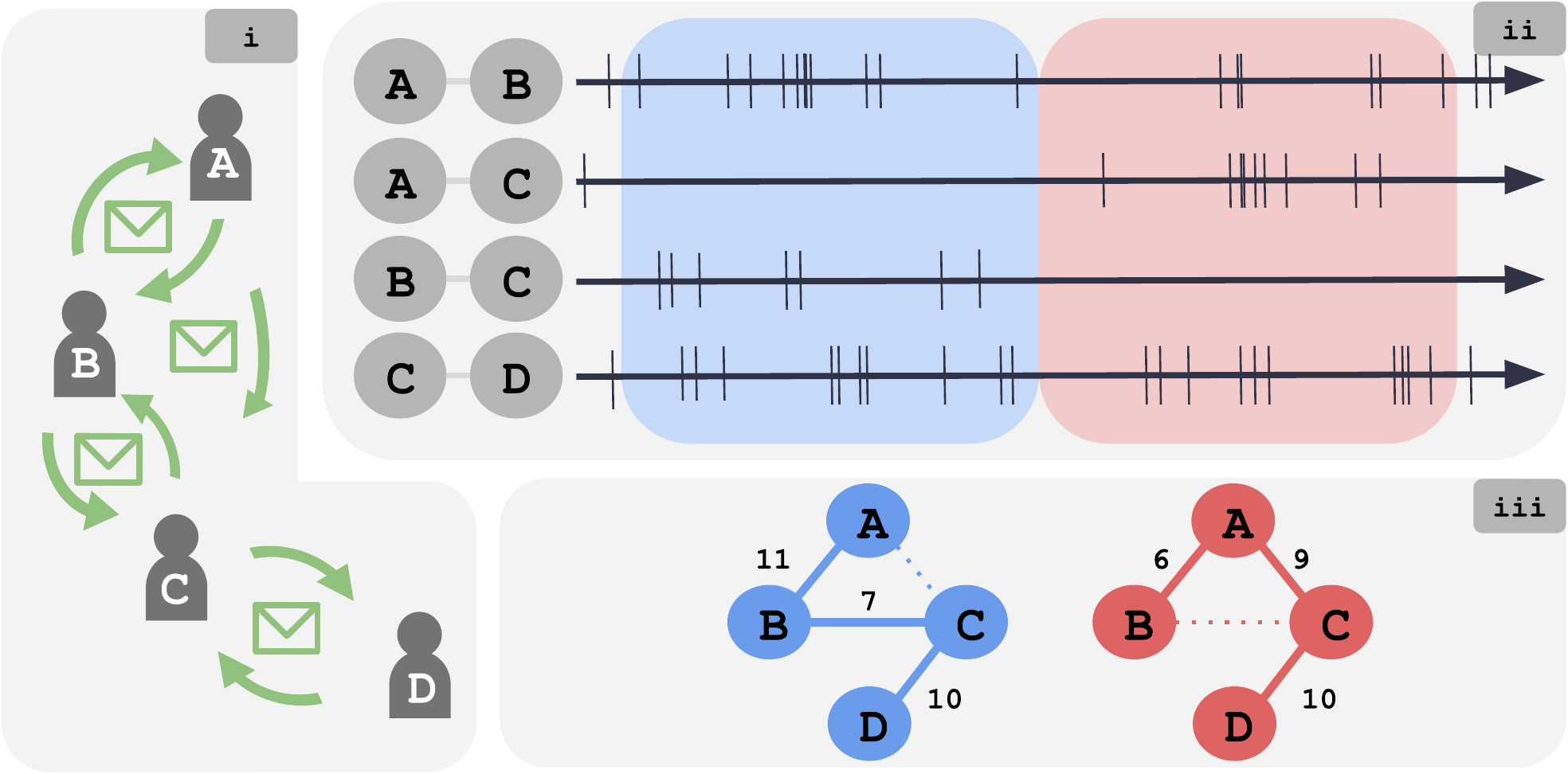}
    \caption{Network reconstruction from recorded interactions. \textit{(i)}, In a simple social system, individuals \textit{A-D} contact each other through directed letters, emails, phone calls, etc; \textit{(ii)}, we model undirected interactions as a time series of contact events, reconstructed from recorded contact metadata (sender, receiver, timestamp). Datasets are constrained to observation periods or snapshots (blue and red). \textit{(iii)} Given an observation period, we can reconstruct a static network by creating links between nodes \textit{A-D} if we observe at least one contact (solid lines). We do not account for a link if we do not observe data during the observation period (dashed lines). A common technique is to assign a link weight $w$ (numbers) based on the number of contacts.}
    \label{fig:fig1}
\end{figure}

% Basics of social networks
Much of the progress in understanding the communication patterns and social behaviour has been made by conceptualising them as patterns in networks ---a way of modelling social systems where individuals (or nodes) interact with each other (via links or edges)~\cite{Rombach2013}---.
Communication metadata has provided crucial access for understating social networks as it allows us to construct exchange networks as proxies for the underlying social system that generated them.  
Figure~\ref{fig:fig1} depicts an example of how interactions can be represented as networks: from a history of communication between two people or nodes, we create links or edges if there has been an interaction, and analyze both reconstructed networks and the time series of interactions.
%A major breakthrough associated to such data is that social networks are far from random: they are highly structured and intimately coupled to human behaviour~\cite{barabasi2002, Saramaki2015, Miritello2013b}. 

%confirming old theories
This approach has confirmed several prominent sociological theories. % which were not originally based on large-scale data. 
Milgram's notion of ``six degrees of separation''~\cite{Milgram1967}, where any two people are expected to be at a path of at most six people away from each other, was popularised in the 60s but only later studied with large-scale data, and is now known as small-world phenomena ~\cite{KarsaiSmall}. 
Similarly, in the 70s, Mark Granovetter hypothesised the "strength of weak ties", which states that the rarely-used weak social links keep networks connected by bridging separate social groups. This was later shown to hold at the level of a country using a communication network of millions of mobile phone users~\cite{Onnela2007,Onnela2009}.

%new theories
These new approaches have also produced completely new theories of social behaviour and the system-level phenomena that emerges from it. 
One of the first key findings was large degree heterogeneity. That is, the number of neighbors varies considerably, where a majority of people have little connections yet a small number of individuals are highly connected. This has major implications on, e.g., disease and information spreading \cite{Kivela2012}. 
%In turn, this phenomenon can be linked to the mechanism of preferential attachment where highly connected individuals are more likely to become even more connected through their previous connections \cite{Barabasi1999}. 
The number of connections is of course limited as there are temporal, spatial and cognitive constraints into maintaining connections, which has lead to common strategies that people follow in managing their communication in the contemporary settings~\cite{Saramaki2014, Heydari2018, MiritelloTimeAllocation}. In this sense, people have robust social signatures: they tend to devote a similar amount of attention to their active circles of contacts, even if the circle of contacts changes~\cite{Saramaki2014, Heydari2018}. 

%new theories continued, timings
Large-scale data has also facilitated the analysis of different dimensions of temporality in human behaviour, including sequential events ~\cite{Barabsi2005,KarsaiBursty} and daily or regular cycles \cite{Aledavood2015a}. Event sequences are often not regular or uniformly random, but the contacts are organised in bursts~\cite{Barabsi2005,KarsaiBursty}. This tells us about the possible processes of how the communication originates from the individuals, such as internal models of priority queues. % For example, internal models like priority queues or self-exiting point processes can be used to explain these bursty patterns.
Despite the small-world phenomena, information does not circulate as freely as burstiness can hinder information flow~\cite{KarsaiBursty, Barabsi2005, KarsaiSmall}. 
% This effect is amplified as the connections at the bridge positions are especially bursty~\cite{Urena2020}. 

The idea that communication mediums have an effect on societal structures and practices has been actively studied from historiographical perspectives. Harold Innis in ''The Bias of Communication'' ~\cite{Innis1951} explored the notion that each historical epoch is distinguished by dominant forms of media % that record and transform information into systems of knowledge 
consonant with the institutional power. Inspired by him, Marshall McLuhan stressed the impact of electronic technology %---as a network of extensions to our own nervous system--- 
on human behaviour and social organization ~\cite{McLuhan1988}. 
%In this context, electronic technology has facilitated the analysis of human behaviour through the analysis of its auto-recorded digital traces. 
%Similarly to these digital traces, the earlier communication left paper traces in forms of letters used as communication medium. 
Social theories that have informed social network analysis have also been used in historical contexts. Lux and Cook~\cite{Lux1998} used the concept of weak ties in scientific exchange in the seventeenth century by contrasting the role of closed circles of academies in England and France and open networks of interactions that facilitated, e.g., the work of Dutch scientists. The small-world phenomena has also been studied on historical networks~\cite{langmead-et-al-2016} in the ''Six Degrees of Francis Bacon'' project~\cite{warren2016six}. 

While letters are not automatically collected into large databases similar to some forms of digital communication, recently there have been large and concentrated efforts to collect and unify such large historical epistolary data  sets distributed across different countries and collections. Metadata about the letters have been aggregated and provided for the research community through web services, such as Europeana\footurl{http://www.europeana.eu}, Kalliope\footurl{http://kalliope.staatsbibliothek-berlin.de}, The Catalogus Epistularum Neerlandicarum\footurl{http://picarta.pica.nl/DB=3.23/}, Electronic Enlightenment\footurl{http://www.e-enlightenment.com}, ePistolarium\footurl{http://ckcc.huygens.knaw.nl/epistolarium/}, the Mapping the Republic of Letters project\footurl{http://republicofletters.stanford.edu}, SKILLNET\footurl{https://skillnet.nl}, correspSearch\footurl{https://correspsearch.net} and the Early Modern Letters Online (EMLO) catalogue\footurl{http://emlo.bodleian.ox.ac.uk} 
%with a particular focus on the Republic of Letters (RofL)
~\cite{van-den-heuvel-2015,van-miert-2016,hotson-wallnig-rrl-2019, Ravenek2017}.  

% What is rofl
%Although contemporary technologies facilitated many of the new findings on human behaviour, such patterns might still be present in historical contexts, and not mere artifacts of current communication channels. 
We contextualize epistolary communication by examining the Republic of Letters (RofL). The RofL was a self-proclaimed community of scholars between the fifteenth and eighteenth centuries that consisted of overlapping networks. In these networks scholars communicated by letters and exchanged knowledge, news, and material objects such as books ~\cite{van-miert-2016}. This group is often described in idealistic terms as a self-regulating open community of scholars not limited by political borders or religious convictions. Nevertheless, these scholars could face persecution on account of religion and politics and often had to choose their words carefully due to the risk that their letters were intercepted, duplicated and even published ~\cite{Daston1991}. Openness, confidentiality and secrecy co-existed and had an impact on the topological nature and development of these networks of scholars in which knowledge exchange was not merely reciprocal, but hierarchical as well ~\cite{van-den-heuvel-2015, vandenHeuvel2016}.
These overlapping networks were in continuous flux. New members were often introduced by central gatekeepers who played a strong regulating and mediating role to keep these networks in balance, but that could also fall apart, resulting in structural holes ~\cite{vanVugt2019}. 

% RofL and contemporary communication
Some characteristics of epistolary communication might be analogous to contemporary forms of communication. The RofL was a community organized around specific social foci that included scholars and the clergy \cite{Feld1981}. In current settings, such focused communication occurs in emails sent within organizations or in specialized internet forums. A major difference between epistolary and contemporary channels is related to immediacy. In epistolary communication, the time between contacts had a considerable time gap constrained by exogenous factors such as distance and logistics, which contrasts with digital channels where waiting times depend mostly on users' behaviour \cite{McLuhan1988}. Now, both historical and contemporary datasets have biases. In the former case, a major source of bias stems from the data-collection process itself  \cite{Ryan2020, Ryan2021}, with researchers slowly building an epistolary archive from different sources; in the latter, auto-recorded data is usually one of many communication channels and temporal windows are shorter. 

Our goal is to investigate how several of the prominent results on contemporary communication hold for letters sent between 1510--1900. We analyse correspSearch, a large-scale epistolary data set representative of the RofL and compare it to four reference data sets on contemporary communication using email, two digital platforms, and mobile phones (see Materials and Methods). We find striking resemblance between historical and contemporary communication patterns, including in the distribution of dyadic temporal features, the robustness of social signatures, and certain network-wide characteristics. We note, however, that there are also considerable structural differences. For these negative results, we can't assure whether they are caused by social phenomena or the biases from the collection process of historical data.

\begin{comment}
\subsection{Republic of Letters: Early Modern Letters Online}

Postal communication in 1500--1800 allowed ordinary men and women to scatter letters across and beyond Europe. This exchange helped knit together % what contemporaries called 
the {\em respublica litteraria}, or Republic of Letters (RofL), a knowledge-based civil society, crucial to that era’s intellectual breakthroughs, and formative of many modern European values and institutions. To enable effective Digital Humanities research on the epistolary data distributed in different countries and collections, metadata about the letters have been aggregated, harmonised, and provided for the research community through the Early Modern Letters Online (EMLO) union catalogue\footurl{http://emlo.bodleian.ox.ac.uk} about the Republic of Letters 1500--1800 ~\cite{hotson-wallnig-rrl-2019}.

\end{comment}

%%%%%%%%%%%%%%%%%%%%%%%%%%%%%%%%%%%%%%%%%%%%%%%%%%%%%%%%%%%%%%%%%%%%%%%%%
\section{Results}

We compare metadata from correspSearch with four modern communication datasets. Each of these datasets consists of logs of interactions including a sender, a receiver and a timestamp. 
We highlight the diversity of communication channels, which include mobile phone calls (\textit{Mobile}), emails (\textit{Email}), Internet community boards (\textit{Forum}) and wall-posting on a social platform (\textit{Platform}) (see Materials and Methods). Taken together, these datasets correspond to several instances and channels of auto-recorded communication, and we claim that these epistolary datasets represent a historical equivalent of contemporary human communication. 

We present findings of common similarities and differences between the historical and contemporary datasets from four analytical approaches: (i) aggregate static networks, (ii) features of pairwise communication (iii) relationship between communication and local topology, and (iv) social signatures of ego networks. 

\subsection{Analysis of the static graph}

We first model communication as aggregate static networks~\cite{Saramaki2015}, where the temporal dimension is projected out by disregarding the contact times. This approach allows us to determine whether there are any major structural similarities between the historical and contemporary networks. As several of the static network features have been studied in historical correspondence datasets \cite{Ryan2020, Ryan2021, Ahnert2015}, we describe these results only briefly, and include a more detailed analysis on Supplementary Information 1. We find that the epistolary network snapshots are similar to other communication networks only to some extent, with the main difference stemming from over-represented nodes. 

The interpretation of static networks differs over large observation periods. For our correspondence data, dyads are constrained to participant's active life, but the overall network can be thought of as the overlapping activity from dyads, where larger structures might not represent, e.g., communities, but temporally overlapped communities, which are more difficult to interpret. We ameliorate some of these issues by aggregating data over different 50-year periods, resulting in a series of static graphs. 

We present a selection of common network statistics: centrality measures, tie overlap and the distribution of shortest paths. Centrality measures are informative of the relative importance  and roles of a node in the network. In this sense, \textit{degree} $k$ measures the number of connections, while \textit{betweenness} $C_B$ is related to the number of shortest paths that pass through a node. Tie overlap $O$ is link-level statistic used to assess whether an edge is contained within overlapping circles of friends, or serves as a "bridge", and the distribution of shortest paths $l$ helps characterize the small-world phenomena. 

\begin{figure}[ht]
    \centering
    \includegraphics[width=.95\textwidth]{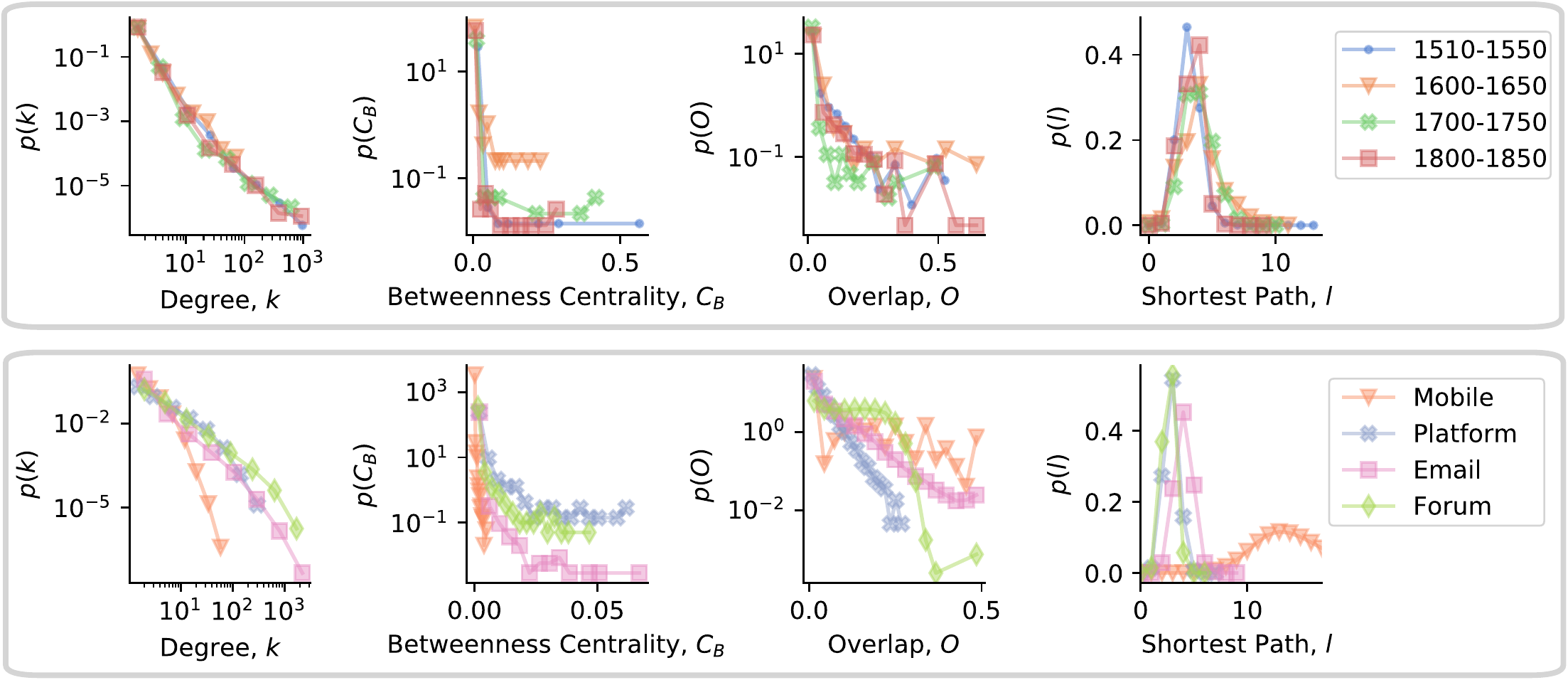}
    \caption{Static features of  communication networks, where the top (bottom) rows contains features of the correspondence (reference) networks. Epistolary datasets are aggregated in 50 year periods. Columns represent distributions of (left to right) degree $k$, betweenness centrality $C_B$, topological overlap $O$ and average shortest path length $l$. For ease of comparison, we do not include individual data points, but present averaged binned values. }
    \label{fig:static}
\end{figure}

On Figure \ref{fig:static} we compare static network statistics for the correspondence and reference datasets. We find similarities in terms of the heavy-tailed degree distribution \cite{Holme2019}, decaying centrality measures and average shortest-path distribution, with epistolary networks exhibiting small-world phenomena even at 50-year aggregation periods. Nevertheless, in correspondence networks some nodes have an outsized impact as compared to most important nodes of the reference networks.
This is reflected in (i) slower decay for the degree distribution, where the tail reflects an upwards flattening curve in the log-log axis; (ii) overall larger centrality measures, e.g., some nodes score $C_B > 0.5$ betweenness, meaning that more than half of all shortest paths go through them as opposed to all other networks having $C_B < 0.08$.  Notably, the overlap distribution might the most affected from sampling, as it requires sampling dyads as opposed to nodes. Th mobile phone network, the data set with most individuals, displays the most differences when compared to the other references, with fast degree and betweenness centrality decay, and longer shortest paths. 

\subsection{Features of human communication}
\begin{figure}[ht]
    \centering
    \includegraphics[width=.8\textwidth]{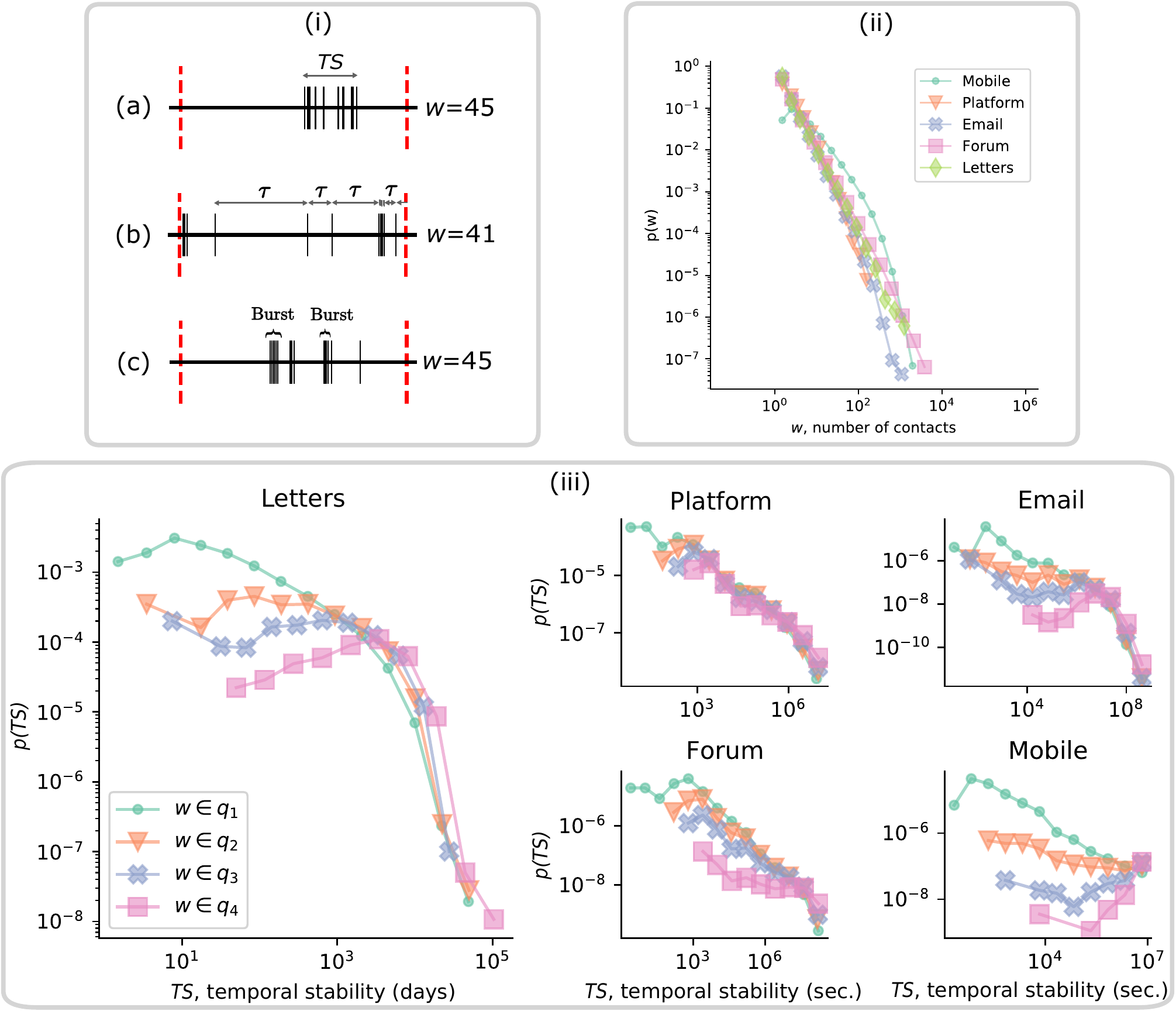}
    \caption{Temporal features of human communication. (i) Visual representation of different temporal features for similar number of contacts, $w$. (\textit{i, a}) Temporal stability $TS$, where we only consider the first and last times during the observation period. (\textit{i, b}) Estimation of the inter-event time (IET) distribution $\tau$, the elapsed time between consecutive events.(\textit{i, c}) Example of bursty behaviour, where bursts of consecutive calls are followed by longer IETs. The burstiness coefficient is defined as $B=\frac{\sigma - \mu}{\sigma + \mu}$ for $\mu$, $\sigma$ the mean and standard deviation of $\tau$. (ii) Empirical distribution of number of contacts $w$ for the epistolary dataset and four reference datasets. (iii) Empirical distribution of temporal stability $TS$ for the epistolary and reference datasets. For each case, we plot the distributions conditioned on link weight quartiles. For ease of comparison, we do not include individual data points, but present averaged binned values.}
    \label{fig:w_tsb}
\end{figure}

We now explore \textit{how} communication occurs between pairs of individuals, where the focus is on different tie-level temporal dynamics, and not only on the network structure.
%We analyze features such as communication intensity, the time between contacts or the duration of the overall contacts.
%%% Notably, in the epistolary datasets the observation period is substantially larger than the lifespan of its individuals, whereas in most modern datasets do not capture changes over the course of an individuals life. This data thus contains examples of individuals and dyads that communicated extensively during larger periods of time. 
Figure~\ref{fig:w_tsb} (\textit{i}) depicts a visual example of approaches to modelling communication sequences, where we obtain statistics from a time series of interactions,
%for similar number of contacts $w$
 such temporal stability $TS$, the inter-event times $\tau$ and burstiness $B$. 
The total contact count $w$ has been used in communication networks as a proxy for tie strength.
Theoretically, Granovetter \cite{Granovetter1973} characterized four dimensions of tie strength: time, emotional intensity, intimacy and reciprocity. In practical terms, however, these dimensions are not straightforward to observe from interaction data, whether because of incompleteness or because of lack of a natural metric or measure~\cite{Carmines1982}.
In the Granovetter sense, however, the proxies for tie strength inform of topological roles of ties, where weaker ties serve as bridges between overlapping circles of friends \cite{Granovetter1973, Urena2020}. 

Figures~\ref{fig:w_tsb} (ii-iii) shows statistics of the number of contacts $w$ and temporal stability $TS$, respectively. 
Aggregate contacts $w$ are highly inhomogeneos for all datasets, displaying heavy-tailed decay. Note that $w$ directly impacts many temporal features. For this reason, we show distributions conditioned on the number of contacts $w$ categorized by quartiles. 
On the other hand, $TS$ displays major inter- and intra-dataset variability. Chiefly, we identify a mix of quartile-independent sharp decay (CorrespSearch, Email), heavy-tailed decay (Platform, Forum) or quartile-dependent growth and decay (CorrespSearch, Email, Forum and Mobile).
These different patterns might arise due to an interplay of the observation period and tie decay. In Mobile we do not expect to observe tie decay since the observation window is short ~\cite{Miritello2013b, Navarro2017}, and thus the $TS$ differs conditioned on $w$ quartiles. Email and Letters, on the other hand, display a mixture of the latter behaviour and sharp decay, constrained by people's active life (within a company, in the case of Email). 

%Now we focus on the time between consecutive contacts, known as the inter-event times (IETs).
The distribution of times between consecutive contacts an individual makes, known as inter-event times (IETs), is an important feature of human communication that has undergone extensive scrutiny in the contemporary communication patterns \cite{Goh2008, KarsaiSmall,KarsaiBursty,KarsaiCorrelated,malmgren2008poissonian}.
In epistolary data, IETs have been analysed for only a handful  of exceptionally prominent individuals,
%IETs have already been analyzed for epistolary data for selected individuals, 
and these finding indicate similarities between contemporary and correspondence data~\cite{Oliveira2005, Malmgren2009}. We expand on this work by focusing on large-scale behaviour and in other statistics. 
Just as with the degree distribution, we find that IETs tend to be very heavy-tailed and bursty~\cite{Goh2008, KarsaiSmall}. This means that the time between events shows high variability, with long resting times followed by short \textit{bursts} of consecutive contacts.

\begin{figure}[ht]
    \centering
    \includegraphics[width=.6\textwidth]{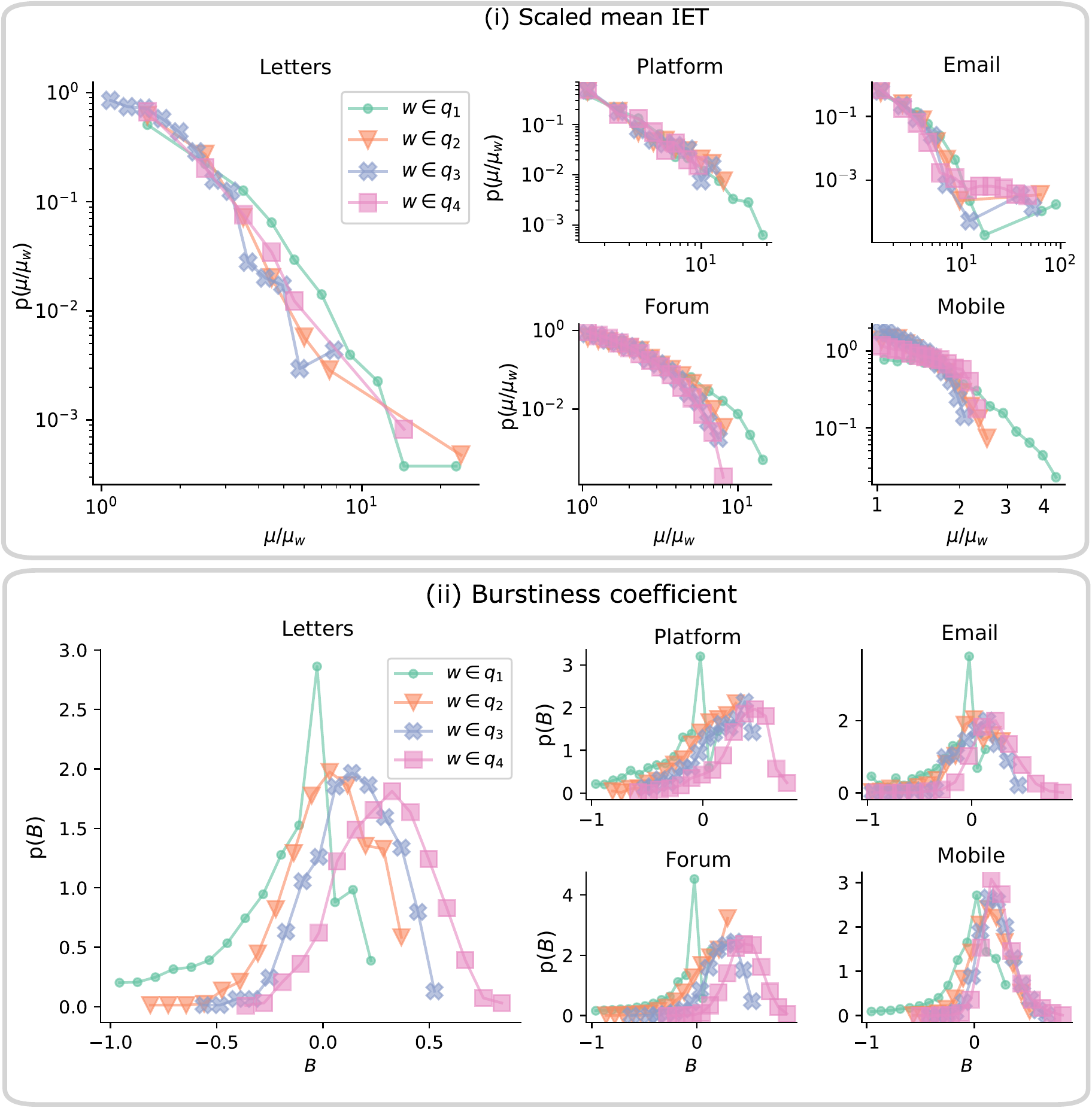}
    \caption{Distribution of different values derived from the IET distribution for correspSearch and our four reference datasets. The scale is days for correspSearch, and seconds for the rest of the datasets. (\textit{i}) Scaled mean IET $\mu / \mu_w$, (\textit{ii}) Burstiness coefficient $B$.}
    \label{fig:iets}
\end{figure}

On Figure \ref{fig:iets} we focus on two statistics of the IET distribution --- mean $\mu$ and burstiness $B$---. %%%%%%%To compute our statistics, we require to have at least three observations per tie. 
%in Figure \ref{fig:iets}(top) we report the average time between two contacts scaled by the average IET given for number of contacts $\mu/\mu_w$. W
We uncouple $\mu$ from the number of contacts by scaling the distribution by $\mu_{w_i}$, the mean distribution value for the quartile category $i$. Previous research has shown that the distributions of $\mu / \mu_{w_i}$ collapse onto a single distribution for all different $w_i$ ~\cite{Goh2008, KarsaiSmall, Saramaki2015}. We replicate this result for most of our reference datasets and for Letters, suggesting that the IET distribution for epistolary data follows a pattern observed before for multiple large contemporary communication data sets~\cite{Goh2008,KarsaiSmall,Candia2008}, and for the set of prominent individuals in historical correspondence \cite{Oliveira2005, Malmgren2009}. We would expect people in close contact to have short average IETs~\cite{Urena2020}; nevertheless, given the large timespan of the epistolary data, the closeness of contacts could change throughout a person's lifetime.

All datasets display larger burstiness values for larger $w$, with the epistolary data following similar average values the Platform and Email datasets ($B$ around -0.19, 0.02, 0.14 for the first three quartiles of Letters, Platform and Email; the last quartiles are 0.28 for Letters, 0.34 for Platform and 0.20 for Email). 
Although burstiness has been explained by behavioural patterns, such as mentally prioritising tasks \cite{Barabsi2005} or cyclic activity periods \cite{malmgren2008poissonian}, some of the bursty character of epistolary data could also be explained by the fact that news sharing and copying played an important role in scholarly correspondences ~\cite{Colavizza2014, Dooley2010}. 

\subsection{Granovetter Effect}

We now examine the Granovetter effect, %the idea strong and weak links have different structural properties on social networks~\cite{Granovetter1973, Granovetter1985}. Briefly, 
the theory states that strong links are characterized by overlapping circles of friends, whereas weak links serve as bridges~\cite{Granovetter1973}. This phenomena has been empirically observed by measuring the strength of ties via contact calls, although it can also be observed for several communication features by uncoupling from $w$~\cite{Urena2020}. 
%Here we focus the Granovetter effect, which links behavioural communication features with static network descriptions. 
In empirical social network analysis, the embeddedness of a tie can be measured via topological overlap~\cite{Onnela2007, Urena2020}. %, which can be regarded as a tie-level clustering coefficient. Given two nodes, overlap measures the percentage of common neighbors out of all neighbors, so that unitary overlap implies that all the neighbors of two nodes are neighbors common, whereas zero overlap implies that there are no common neighbors between the two nodes. 

%For $i, j \in V$ , and $\mathcal{N}(i)$ the set of neighbors of node $i$, we define topological overlap as the Jaccard similarity between the sets of neighbors %Change and paraphrase

%\begin{equation}
%    O_{ij} = \frac{|\mathcal{N}(i)\cap \mathcal{N}(j)|}{|\mathcal{N}(i) \cup \mathcal{N}(j)|}
%\end{equation} 

\begin{figure}[ht]
    \centering
    \includegraphics[width=.9\textwidth]{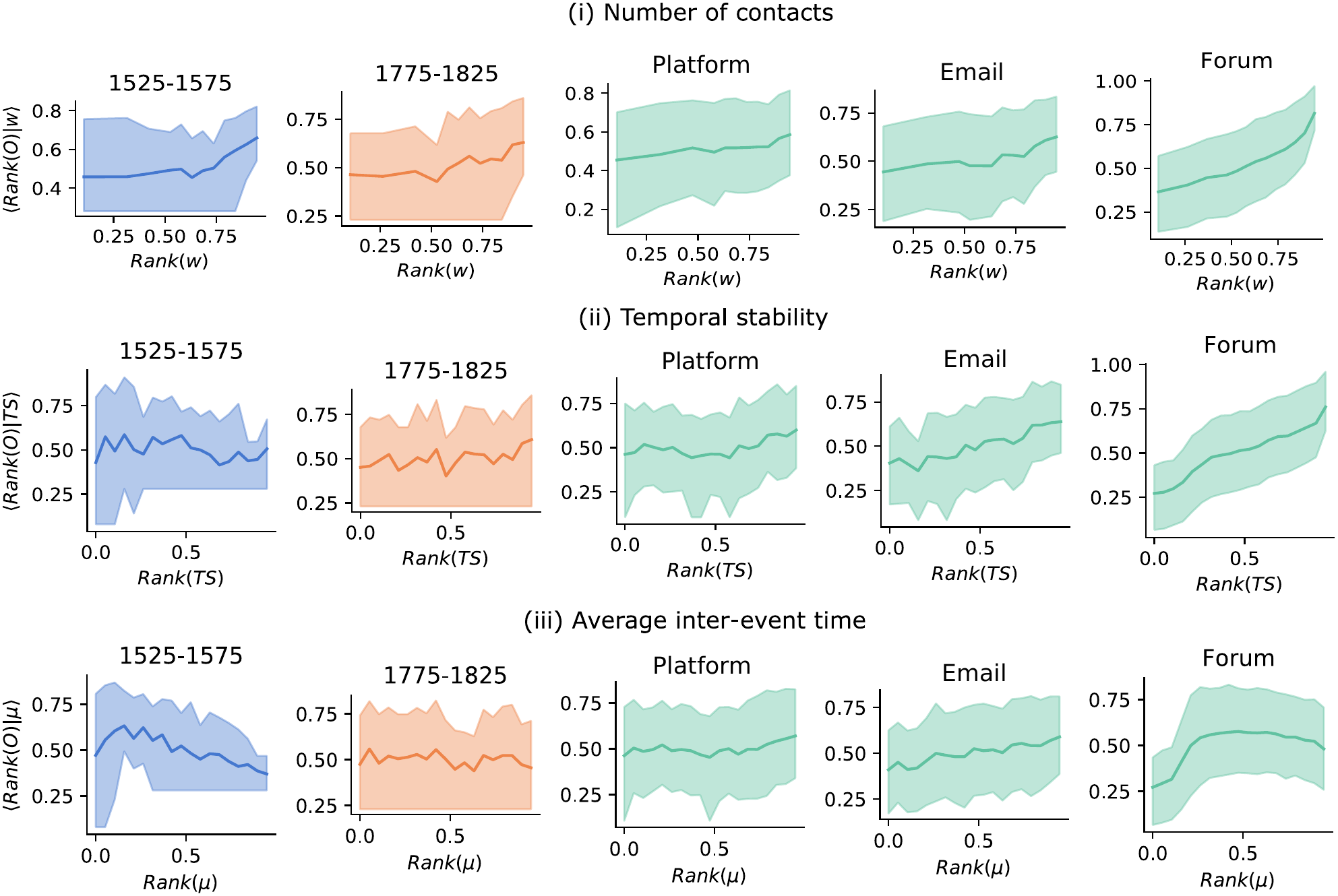}
    \caption{Granovetter effect for different features of communication on (\textit{left} to \textit{right}) two snapshots of correspSeach (1525-1575) and reference datasets Platform, Email and Forum (the effect has already been descibed in Mobile by \cite{Urena2020}). The embeddedness of ties, as measured by topological overlap, changes in association with behavioural communication features, although not always in a linear fashion. Shaded area represents 80\% of the distribution. Features of communication: (\textit{i}) number of contacts $w$, (\textit{ii}) temporal stability $TS$ and (\textit{iii}) Mean IET $\mu$. The Granovetter effect is most visible for $w$. Using $TS$ as a proxy for tie strength we do not observe the effect for correspSearch and we only observe it partly for Platform. For $\mu$, the effect (decrease of overlap with larger average IET) is visible for the period 1525-1575 and partially for larger values of Forum. }
    \label{fig:granovetter}
\end{figure}

We explore the Granovetter effect on Figure~\ref{fig:granovetter} by comparing overlap with $w$, $TS$ and $\mu$. We find an association between some of the features and overlap, with a visible effect for all datasets using $w$. This indicates that correspondence data exhibits a salient feature of social networks, where user behaviour tends to be found within local community structures. Notably, for $TS$ we did not find a clear association for the epistolary data, as well as for $\mu$ in the period 1775-1825. This might be related to sampling biases in the network and in the time-series. In the first case, overlap is a measure based on triads, which tend to be less likely to be sampled. Second, tie-level dynamics might be affected by irregular sampling at different time periods. This could explain why we observe the effect for $\mu$ in 1525-1575 but not in 1775-1825. 

%The idea that weak and strong ties are related to network topology can also be explored via a percolation analysis. [TODO: Add Percolation]
%\newpage
\subsection{Social Signatures}

Ego networks, where the focus is on central nodes (egos) and their immediate neighbors (alters) serve as useful characterizations of user behaviour. %As we have seen, some individuals are over-represented in the epistolary datasets %, while a majority ($\approx 78\%$) only has one neighbor. This 
%an imbalance that motivates analyzing the ego networks of over-represented individuals as more complete samples. 
We focus on social signatures, the idea that people divide their attention to their high-ranking contacts in a persistent fashion across time and uniquely to each ego~\cite{Saramaki2014, Heydari2018}. This does not mean that the alters don't change --- they are highly dynamic---, but the attention devoted to high-ranking alters is persistent. We inspect the ego networks of different people at consecutive snapshots, where on each snapshot we create an ego network with the number of outgoing contacts as link weights, where we focus only on outgoing contacts to characterize the behaviour of the ego. Afterwards, we build the social signature by ranking the alters and obtaining the fraction of outgoing contacts. 

In the seminal work where social signatures were observed for the first time \cite{Saramaki2014},
Saramäki et al. divided the observation period into three snapshots, thus obtaining three social signatures for each individual. They compared social signatures for different intervals using the Jensen-Shannon Divergence (Signature Distance $d$), a measure for comparing probability distributions, and for reference they also compared social signatures to other individuals. To adapt their analysis to our dataset, we incorporate two filters regarding the number of alters during an interval, and the length of the interval. We determine social signatures on a interval of size $b$, measured in years for Letters, Email and Forum, and in months for Platform and Mobile. Following~\cite{Saramaki2014}, we compare social signatures for a given individual in $m=3$ adjacent time intervals, and we compare signatures of other individuals during time intervals of the same length. We determine a limit on the minimum number of out-neighbors $n_{o-ng}$ that an ego must have during all intervals in order to be a valid social signature, and we vary this parameter for a range of values. 

\begin{figure}[ht]
    \centering
    \includegraphics[width=.8\textwidth]{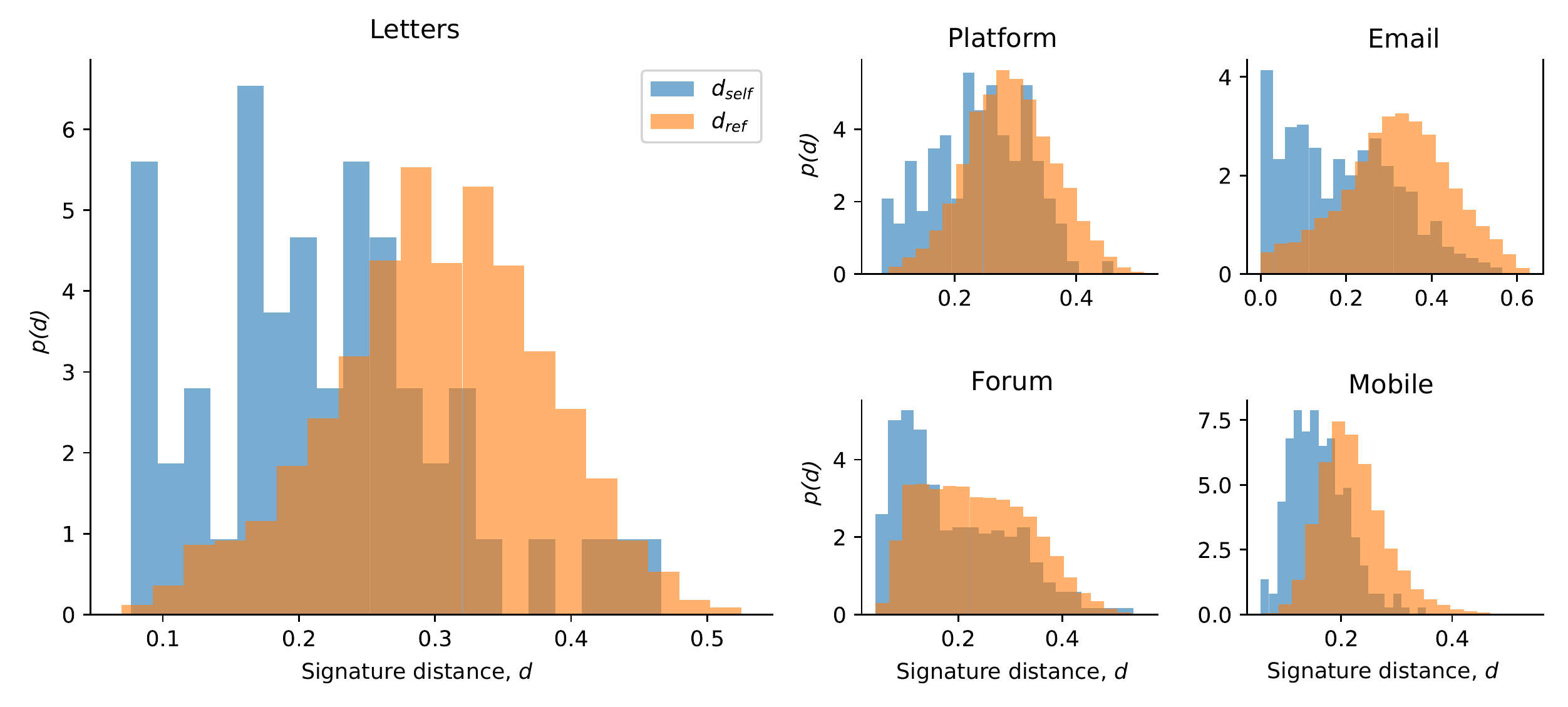}
    \caption{Comparison between self-distances $d_{self}$ and reference distances $d_{ref}$ for ego networks of all datasets. The average self distances are smaller than average reference distances for all parameter combinations, implying that social signatures tend to be more similar for the same person during different times, than to other people's signatures. We performed a two-sided T-test for difference in means, finding that for all datasets the mean $d_{self}$ differed from $d_{ref}$ with all estimated p-vales $<10^{-13}$. The estimated means for all datasets are: Letters $\hat{d}_{self}=0.218 \pm 0.048$ and $\hat{d}_{ref}=0.302 \pm 0.008$; Platform $\hat{d}_{self}=0.241 \pm 0.025$ and $\hat{d}_{ref}=0.292 \pm 0.002$; Email $\hat{d}_{self}=0.191 \pm 0.019$ and $\hat{d}_{ref}=0.317 \pm 0.0001$; Forum $\hat{d}_{self}=0.189 \pm 0.019$ and $\hat{d}_{ref}=0.235 \pm 0.001$;  Mobile $\hat{d}_{self}=0.162 \pm 0.014$ and $\hat{d}_{ref}=0.227 \pm 0   .002$. Confidence intervals for the mean reported at 95\%. }
    \label{fig:soc_signs}
\end{figure}

Figure~\ref{fig:soc_signs} depicts the results for our analysis of social signatures for filter values $n_{o-ng}=5$, and different bin sizes $b$ depending on the dataset; we were, however, able to replicate the results of~\cite{Saramaki2014, Heydari2018} for all datasets and for all filter values in Letters  (see Supplementary Material 2). This suggests that individualized patterns of communication that are consistent are not a feature of contemporary communication networks, but were also persistent in older and less immediate forms of communication as well. Even more, our results hold for observation windows ranging from 1 to 10 years, and totaling comparisons ranging 3 to 30 years considering three observation windows. This indicates that these individual patterns can be observed not only in shorter timescales, but also in periods spanning a large proportion of an adult life. 

\section{Discussion}

We used common methods for analyzing social networks and applied them to a large epistolary dataset in an attempt to uncover whether major differences exist between contemporary and historical communication networks. For reference, we used four contemporary datasets of communication data. Our results strongly suggest that prominent characteristics of contemporary communication networks are also found in historical contexts, validating that these characteristics are related to fundamental human behavior, not only to specific communication mediums practices. Despite these major positive results, we also found differences in the contemporary and historical communication patterns. 
We might attribute these negative results to differences in the completeness of the datasets, since we depend on letters that have survived and been included in collections. We note, however, that they might also stem from differences in the use of media and communication practices.

We divided our analyses in four main parts: global network structure and characteristics, dyadic communication patterns, their relationship to local topology, and analysis of ego networks. Regarding global structures, we found the epistolary datasets to be both highly-local and deeply connected---in the sense that most nodes have a very small number of connections, while still maintaining relatively short average shortest path distributions. However, we also found that a majority of nodes are under-represented in the network, having less-than-expected connections. Indeed, since epistolary datasets tend to focus on relevant historical figures, we centre the rest of our analysis on dyadic interactions and ego networks. 

Dyadic interactions display major similarities both in terms of the number of contacts and in the distribution of inter-event times, even when considering that the latter follows naturally different scales. Our results suggest that major features of dyadic communication can be described with known mechanisms of average IET decay, and that the overall burstiness distribution is in line with contemporary communication channels. This is slightly striking, as one could expect that limitations in using letters, such as less immediacy, could imply greater regularity in communication. Further analyses, however, could reveal how burstiness varies across several years. 

We also found some evidence that epistolary networks also reflect the Granovetter effect, where stronger ties are contained within overlapping circles of friends, while weaker serve as bridges. 
To some extent, our most interesting results were related to ego networks, possibly due to more control in filtering come complete samples of ego networks. In this case, we found that individuals in previous centuries communicated in a similar manner than we do today: uniquely to individuals and persistent in time. 

\subsubsection{Acknowledgements}

We thank Dirk van Miert, and Sebastiaan Derks for discussions and an the possiblity of using ePistolarium data in our research. Thanks to
Sebastian Dumont for providing the correspSearch data and discussions.
Our work is continuation of the collaborations in the EU COST action Reassembling the Republic of Letters and with University of Oxford in the Cultures of Knowledge project and the Early Modern Letters Online service. Funding for the current work was provided by the Academy of Finland and the EU project In/Tangible European Heritage (InTaVia)\footurl{https://intavia.eu}.

\section{Materials and Methods}

%%%%%%%%%%%%%%%%%%%%%%%%%%%%%%%%%%%
\subsection{Data}

%\todo{Data model (Petri, Jouni)}

%\todo{Sampo Model and Sampo-UI (Eero)}

We include four datasets of contemporary human communication that encompass different communication channels and environments, including email data, mobile calls and online postings. More specifically: 

\begin{table}[ht] 
    \caption{Datasets analyzed and compared in this paper}
    \centering
    \label{tab:datasets}
        \begin{tabular} {| p{0.1\textwidth} |  p{0.58\textwidth} |  p{0.2\textwidth} | p{0.07\textwidth} |} %{| l |  l |  l |}
        \hline
        \textbf{Dataset} & \textbf{Content} & \textbf{Observation Period}  &\textbf{Source} \\
         \hline
%        RofL     & Epistolary data of ePistolarium\footurl{http://ckcc.huygens.knaw.nl/?page_id=43 }, an aggregated collection of ca \num{20 000}  Dutch correspondences~\cite{van-den-heuvel-2015,van-miert-2016} related to the Republic of Letters~\cite{van-miert-2016,hotson-wallnig-rrl-2019}\\
%        \hline
%        correspSearch & Epistolary data 1510--1991 of \num{135 000} letters aggregated by the correspSearch project\footurl{https://correspsearch.net} at the Berlin Brandenburg Academy of Sciences and Humanities~\cite{dumont2016}\\
%        \hline
        Platform  &  Wall-posting on Facebook in New Orleans, where a directed link is created if a user posts on another user's wall. Data obtained from crawling public profiles in January 2009, each link tracks activity for one year after the first communication event. Total of $\sim 41$K nodes, $\sim 183.5$K edges and $\sim854.6$K wall posts.  & 1 year  & ~\cite{Viswanath2009}\\
        %\footnote{Data available on: http://socialnetworks.mpi-sws.org/data-wosn2009.html}
        \hline
        Email & Dataset of emails between employees of Enron Corporation, obtained as part of an investigation into the Corporation's collapse. Sampling might focus on 150 users. Total of $\sim 20$K users, $297.5$K edges and $\sim 1.1$M emails.  & 1998-2004 & ~\cite{nraaai15}\\
        \hline
        Forum & Activity on a movie forum, where a contact is created if a user comments on a thread posted by another user. Total of $\sim 6$K users, $\sim 138.1$K edges and $\sim 1.4$M thread comments. & 7 years & ~\cite{Karimi2014}\\
        \hline
        Mobile & Call Detail Records of mobile phone calls from an operator in a European country in 2007, where the operator had approximately 20\% market share. We used a reduced version of the full dataset, where we focused on a region of the country. The small version of the dataset has $\sim 174$K nodes, $\sim 190$K edges and $\sim 6.8$M phone calls. & 4 months & ~\cite{Onnela2007} \\ \hline
    \end{tabular}
\end{table}

The Letters datasets comes from the aggregated data service correspSearch~\cite{dumont2016} provided by the Berlin Brandenburg Academy of Sciences and Humanities. This dataset includes metadata about ca. 135,000 letters sent in 1510--1900.

All data we used had been anonymized from source for all reference datasets, and for the purpose of our research, we also used an anonymized version of Letters. These datasets provide fruitful insights into modern human communication. \textit{Mobile} and \textit{Platform} contains data a large user database, allowing us to examine communication patterns at regional scales in both European and North American contexts. The sampling of both datasets is foci-independent in the sense that it does not follow specific societal units within their region. \textit{Platform}, however, it still represents a sample of people with public Facebook profiles. On the other hand, \textit{Email} and \textit{Forum} contains information from specific contexts (a company and an online community), and thus their interactions might follow certain structures. In the case of \textit{Email}, we might expect that the sampling methodology that favors certain users. The Letters dataset constitutes a unique example of human communication encompassing five centuries of communication data across different countries and time periods. This presents both unique opportunities and challenges, as some of the common assumptions and techniques used for communication networks might not hold true. Notably, these datasets include letters from high-ranking personalities of religion, politics and science from several European countries, and constitutes a (non-random) sample from a communication network spanning five centuries. 

The epistolary datasets have been transformed into Linked Data and published on the Linked Data Finland platform\footurl{https://ldf.fi}~\cite{hyvonen-et-al-ldf-2014} according to the Linked Data publishing principles and other best practices of W3C~\cite{heath-bizer-2011}, including, e.g., content negotiation and provision of a SPARQL endpoint. %\footnote{The homepage of the data service including, e.g., documentation of the data and pointers for linked data browsing and the SPARQL endpoint, is available at: \url{https://www.ldf.fi/dataset/XXX}}. 
However, the data are at the moment not publicly available on the Web due to copyright issues under negotiation with the owners of the primary metadata.

\subsection{Static Network Construction}

We construct static networks from communication data following the commonly-used process described by~\cite{Saramaki2015} and~\cite{Onnela2007}. We process data in the form of logs, which in its most basic form include information of a sender, a receiver and a timestamp. This holds both for the epistolary datasets and the modern reference data. We create an undirected edge between nodes if we have observed at least one letter between, but we keep track of the time series of interactions for further analysis. 

\subsection{Temporal features}

We derive different statistics based on the distribution of IETs. For a sequence of interaction times of at least length three, $\{t_0, t_1, \ldots, t_n\}$, we obtain the $k$th inter-event time $\tau_k = t_{k+1} - t_k$, and estimate moments from this distribution from the empirical observations $\{\tau_k\}$. We use the Kaplan-Meier estimator~\cite{Kivela2015}, which considers  \textit{censored} data, as a means to correct for lack of data in epistolary contexts. We thus obtain $\mu(\tau)$ (in the text $\mu$) and $\tau_{\sigma}$ , the mean and standard deviation of our IETs. We use the burstiness coefficient as proposed by~\cite{Goh2008}, which is defined as $B=(\tau_{\sigma} - \tau) / (\tau_{\sigma} + \tau)$. We measure temporal stability in terms of the first and last observed interactions, $TS = t_n - t_0$.

For most temporal features, we plot distributions conditioned on quartiles of $w$. Given the heterogeneity of $w$ it is not straightforward to partition data into bins of equal size. In Table \ref{tab2:quartiles} we show the composition of quartiles for each dataset. Notably, only Mobile is partitioned into bins of roughly equal sizes (as percentage), for all other datasets, $q_1$ is disproportionately large.

\begin{table}[ht]
\begin{tabular}{cccc|cc|cc|cc|}
\multicolumn{2}{c}{} &
  \multicolumn{2}{c}{$q_1$} &
  \multicolumn{2}{c}{$q_2$} &
  \multicolumn{2}{c}{$q_3$} &
  \multicolumn{2}{c}{$q_4$} \\
\multicolumn{1}{|c|}{Dataset} & \multicolumn{1}{l|}{$\%w <= 2$} & Contacts & Percentage & Contacts & Percentage & Contacts & Percentage & Contacts & Percentage \\ \hline
\multicolumn{1}{|c|}{CorrespSearch} &
  \multicolumn{1}{c|}{0.709} &
  3-4 &
  0.477 &
  5-7 &
  0.162 &
  8-16 &
  0.17 &
  \textgreater{}16 &
  0.19 \\
\multicolumn{1}{|c|}{Platform} &
  \multicolumn{1}{c|}{0.578} &
  3 &
  0.425 &
  4-5 &
  0.208 &
  6-9 &
  0.169 &
  \textgreater{}9 &
  0.20 \\
\multicolumn{1}{|c|}{Email} &
  \multicolumn{1}{c|}{0.734} &
  3 &
  0.434 &
  4-5 &
  0.182 &
  6-10 &
  0.185 &
  \textgreater{}10 &
  0.198 \\
\multicolumn{1}{|c|}{Forum} &
  \multicolumn{1}{c|}{0.649} &
  3 &
  0.358 &
  4-5 &
  0.154 &
  6-14 &
  0.243 &
  \textgreater{}14 &
  0.245 \\
\multicolumn{1}{|c|}{Mobile} &
  \multicolumn{1}{c|}{0.146} &
  3-5 &
  0.28 &
  6-13 &
  0.235 &
  14-39 &
  0.237 &
  \textgreater{}40 &
  0.248 \\ \hline
\end{tabular}
\caption{Quartile composition for each dataset. The first numerical column contains the fraction of linksn that have either one or two contacts, and which were thus not used to compute temporal features. The latter columns contain the composition of each quartile, including their ranges of contacts and the percentage of links in that category.}

\label{tab2:quartiles}
\end{table}

\subsection{Granovetter effect}

For the Granovetter effect, we measure the overlapping circles of friends between an edges via topological overlap. For $i, j$ two nodes, and $\mathcal{N}(i)$ the set of neighbors of node $i$, the topological overlap is the Jaccard similarity between the sets of neighbors %Change and paraphrase
    
\begin{equation}
    O_{ij} = \frac{|\mathcal{N}(i)\cap \mathcal{N}(j)|}{|\mathcal{N}(i) \cup \mathcal{N}(j)|}
\end{equation} 

\section{List of Abbreviations}
\begin{itemize}
\item RofL: Republic of Letters
\item EMLO: Early Modern Letters Online
\item IET: Inter-event Time
\end{itemize}

\subsection*{Authorship}
J.U.C., M.K. and E.H. conceived the main ideas. J.U.C. and M.K. designed the analysis, and J.U.C. performed the analysis. P.L., J.T. and J.U.C collected the data and contributed with analytic tools. C.V.D.H. contextualized historical data and designed analysis. J.U.C. took the lead in writing the manuscript, to which all authors contributed. 

\subsection*{Data Availability}
All datasets except Mobile are available online. Mobile is not publicly available due to a signed non-disclosure agreement, as it contains sensitive information of the subscribers. 
The epistolary dataset is being integrated in the LetterSampo system\cite{hyvonen-et-al-lettersampo-2021,leskinen-et-al-lettersampo-2021}, underway for publishing and using epistolary linked data on the Semantic Web, a new member in the Sampo series of Linked Open Data services and semantic portals\footurl{https://seco.cs.aalto.fi/applications/sampo/}~\cite{hyvonen-sampos-dhn-2020}.

\subsection*{Code Availability}

A selection of the  networks analytic tools discussed in this paper are being integrated in the LetterSampo system.
Code is available at \url{https://version.aalto.fi/gitlab/urenaj1/rofl_communication_patterns}.

% Other catalogue
% http://emlo-portal.bodleian.ox.ac.uk/collections/?catalogue=bodleian-card-catalogue
%%%%%%%%%%%%%%%%%%%%%%%%%%%%%%%%%%%%%%%%%%%%%%%%%%%%%%%%%%%%%%%%%%%%%%%%%%%%

%%%%%%%%%%%%%%%%%%%%%%%%%%%

\bibliographystyle{biolett} %ieeetr}
\bibliography{references}

\begin{thebibliography}{10}
\expandafter\ifx\csname urlstyle\endcsname\relax
  \providecommand{\doi}[1]{doi:\discretionary{}{}{}#1}\else
  \providecommand{\doi}{doi:\discretionary{}{}{}\begingroup
  \urlstyle{rm}\Url}\fi

\bibitem{lazer2020computational}
Lazer DM, Pentland A, Watts DJ, Aral S, Athey S, Contractor N, Freelon D,
  Gonzalez-Bailon S, King G, Margetts H, \emph{et~al.} 2020 Computational
  social science: Obstacles and opportunities.
\newblock \emph{Science} \textbf{369}, 6507, 1060--1062.

\bibitem{lazer2021meaningful}
Lazer D, Hargittai E, Freelon D, Gonzalez-Bailon S, Munger K, Ognyanova K,
  Radford J. 2021 Meaningful measures of human society in the twenty-first
  century.
\newblock \emph{Nature} pp. 1--8.

\bibitem{Onnela2007}
Onnela JP, Saram\"{a}ki J, Hyv\"{o}nen J, Szab{\'{o}} G, Lazer D, Kaski K,
  Kert{\'{e}}sz J, Barab{\'{a}}si AL. 2007 Structure and tie strengths in
  mobile communication networks.
\newblock \emph{Proceedings of the National Academy of Sciences} \textbf{104},
  18, 7332--7336.
\newblock (\doi{10.1073/pnas.0610245104}).

\bibitem{Candia2008}
Candia J, Gonz{\'{a}}lez MC, Wang P, Schoenharl T, Madey G, Barab{\'{a}}si AL.
  2008 Uncovering individual and collective human dynamics from mobile phone
  records.
\newblock \emph{Journal of Physics A: Mathematical and Theoretical}
  \textbf{41}, 22, 224015.
\newblock (\doi{10.1088/1751-8113/41/22/224015}).

\bibitem{Borgatti2009}
Borgatti SP, Mehra A, Brass DJ, Labianca G. 2009 Network analysis in the social
  sciences.
\newblock \emph{Science} \textbf{323}, 5916, 892--895.
\newblock (\doi{10.f1126/science.1165821}).

\bibitem{JariTemp}
Holme P, Saram\"{a}ki J. 2012 Temporal networks.
\newblock \emph{Physics Reports} \textbf{519}, 3, 97--125.
\newblock (\doi{10.1016/j.physrep.2012.03.001}).

\bibitem{Miritello2013b}
Miritello G. 2013 \emph{Temporal Patterns of Communication in Social Networks}.
\newblock Springer International Publishing.
\newblock (\doi{10.1007/978-3-319-00110-4}).

\bibitem{Miritello2012a}
Miritello G, Lara R, Cebrian M, Moro E. 2013 Limited communication capacity
  unveils strategies for human interaction.
\newblock \emph{Scientific Reports} \textbf{3}, 1.
\newblock (\doi{10.1038/srep01950}).

\bibitem{MiritelloTimeAllocation}
Miritello G, Lara R, Moro E. 2013 Time allocation in social networks:
  Correlation between social structure and human communication dynamics.
\newblock In: \emph{Understanding Complex Systems}, pp. 175--190. Springer
  Berlin Heidelberg.
\newblock (\doi{10.1007/978-3-642-36461-7\textunderscore9}).

\bibitem{Barabsi2002b}
Barab{\'{a}}si A, Jeong H, N{\'{e}}da Z, Ravasz E, Schubert A, Vicsek T. 2002
  Evolution of the social network of scientific collaborations.
\newblock \emph{Physica A: Statistical Mechanics and its Applications}
  \textbf{311}, 3-4, 590--614.
\newblock (\doi{10.1016/s0378-4371(02)00736-7}).

\bibitem{Innis1951}
Innis H. 1951 \emph{The Bias of Communication}.
\newblock University of Toronto Press.

\bibitem{McLuhan1988}
McLuhan M. 1988 The role of new media in social change.
\newblock In: G S, F M (eds.), \emph{Marshall McLuhan: The man and his
  message}, pp. 34--35. Fulcrum Inc.

\bibitem{Rombach2013}
Rombach MP, Porter MA, Fowler JH, Mucha PJ. 2013 Core-periphery structure in
  networks.
\newblock \emph{{SSRN} Electronic Journal} (\doi{10.2139/ssrn.2002684}).

\bibitem{Milgram1967}
Milgram S. 1967 The small-world problem.
\newblock \emph{Psychology Today} \textbf{1}, 1.

\bibitem{KarsaiSmall}
Karsai M, Kivel\"{a} M, Pan RK, Kaski K, Kert{\'{e}}sz J, Barab{\'{a}}si AL,
  Saram\"{a}ki J. 2011 Small but slow world: How network topology and
  burstiness slow down spreading.
\newblock \emph{Physical Review E} \textbf{83}, 2.
\newblock (\doi{10.1103/physreve.83.025102}).

\bibitem{Onnela2009}
Onnela JP, Saram{\"a}ki J, Hyv{\"o}nen J, Szab{\'o} G, {Argollo de Menezes} M,
  Kaski K, Barab{\'a}si AL, Kert{\'e}sz J. 2007 Analysis of a large-scale
  weighted network of one-to-one human communications.
\newblock \emph{New Journal of Physics} \textbf{9}, 179.

\bibitem{Kivela2012}
Kivel\"{a} M, Pan RK, Kaski K, Kert{\'{e}}sz J, Saram\"{a}ki J, Karsai M. 2012
  Multiscale analysis of spreading in a large communication network.
\newblock \emph{Journal of Statistical Mechanics: Theory and Experiment}
  \textbf{2012}, 03, P03005.
\newblock (\doi{10.1088/1742-5468/2012/03/p03005}).

\bibitem{Saramaki2014}
Saramäki J, Leicht EA, Lopez E, Roberts SGB, Reed-Tsochas F, Dunbar RIM. 2014
  Persistence of social signatures in human communication.
\newblock \emph{Proceedings of the National Academy of Sciences} \textbf{111},
  3, 942--947.
\newblock (\doi{10.1073/pnas.1308540110}).

\bibitem{Heydari2018}
Heydari S, Roberts SG, Dunbar RIM, Saram\"{a}ki J. 2018 Multichannel social
  signatures and persistent features of ego networks.
\newblock \emph{Applied Network Science} \textbf{3}, 1.
\newblock (\doi{10.1007/s41109-018-0065-4}).

\bibitem{Barabsi2005}
Barab{\'{a}}si AL. 2005 The origin of bursts and heavy tails in human dynamics.
\newblock \emph{Nature} \textbf{435}, 7039, 207--211.
\newblock (\doi{10.1038/nature03459}).

\bibitem{KarsaiBursty}
Karsai M, Kaski K, Barab{\'{a}}si AL, Kert{\'{e}}sz J. 2012 Universal features
  of correlated bursty behaviour.
\newblock \emph{Scientific Reports} \textbf{2}, 1.
\newblock (\doi{10.1038/srep00397}).

\bibitem{Aledavood2015a}
Aledavood T, Lehmann S, Saramäki J. 2015 Digital daily cycles of individuals.
\newblock \emph{Frontiers in Physics} \textbf{3}, 73.
\newblock (\doi{10.3389/fphy.2015.00073}).

\bibitem{Lux1998}
Lux DS, Cook HJ. 1998 Closed circles or open networks?: Communicating at a
  distance during the scientific revolution.
\newblock \emph{History of Science} \textbf{36}, 2, 179--211.
\newblock (\doi{10.1177/007327539803600203}).

\bibitem{langmead-et-al-2016}
Langmead A, Otis J, Warren C, Weingart S, Zilinski L. 2016 {Towards
  Interoperable Network Ontologies for the Digital Humanities}.
\newblock \emph{Int. J. of Humanities and Arts Computing} \textbf{10}.
\newblock (\doi{http://dx.doi.org/10.3366/ijhac.2016.0157}).

\bibitem{warren2016six}
Warren CN, Shore D, Otis J, Wang L, Finegold M, Shalizi C. 2016 {Six Degrees of
  Francis Bacon: A Statistical Method for Reconstructing Large Historical
  Social Networks.}
\newblock \emph{DHQ: Digital Humanities Quarterly} \textbf{10}, 3.

\bibitem{van-den-heuvel-2015}
van~den Heuvel C. 2015 Mapping knowledge exchange in {Early Modern Europe}:
  Intellectual and technological geographies and network representations.
\newblock \emph{International Journal of Humanities and Arts Computing}
  \textbf{9}, 1, 95--114.
\newblock (\doi{10.3366/ijhac.2015.0140}).

\bibitem{van-miert-2016}
van Miert D. 2016 What was the {Republic of Letters}? {A} brief introduction to
  a long history (1417--2008).
\newblock \emph{Groniek} \textbf{204/205}, 269--287.

\bibitem{hotson-wallnig-rrl-2019}
Hotson H, Wallnig T (eds.). 2019 \emph{Reassembling the Republic of Letters in
  the Digital Age}.
\newblock Göttingen University Press.

\bibitem{Ravenek2017}
Ravenek W, {van den Heuvel} C, Gerritsen G. 2017 \emph{The ePistolarium:
  Origins and Techniques}, pp. 309--316.
\newblock London: Ubiquity Press Limited.
\newblock (\doi{10.5334/bbi.26}).

\bibitem{Daston1991}
Daston L. 1991 The {I}deal and {R}eality of the {R}epublic of {L}etters in the
  {E}nlightenment.
\newblock \emph{Science in Context} \textbf{4}, 2, 367–386.
\newblock (\doi{10.1017/S0269889700001010}).

\bibitem{vandenHeuvel2016}
van~den Heuvel C, Weingart SB, Spelt N, Nellen H. 2016 Circles of confidence in
  correspondence.
\newblock \emph{Nuncius} \textbf{31}, 1, 78--106.
\newblock (\doi{10.1163/18253911-03101002}).

\bibitem{vanVugt2019}
van Vugt I. 2019 \emph{The structure and dynamics of scholarly networks between
  the Dutch Republic and the Grand Duchy of Tuscany in the 17th Century}.
\newblock Universiteit van Amsterdam.

\bibitem{Feld1981}
Feld SL. 1981 The focused organization of social ties.
\newblock \emph{American Journal of Sociology} \textbf{86}, 5, 1015--1035.
\newblock (\doi{10.1086/227352}).

\bibitem{Ryan2020}
Ryan Y, Ahnert S, Ahnert R. 2020 Networking archives: Quantitative history and
  the contingent archive.
\newblock In: \emph{Proceedings of the Workshop on Computational Humanities
  Research (CHR 2020)}, Vol-2723.

\bibitem{Ryan2021}
Ryan YC, Ahnert SE. 2021 The measure of the archive: The ro{\-}bustness of
  network analysis in early modern correspondence.
\newblock \emph{Journal of Cultural Analytics} (\doi{10.22148/001c.25943}).

\bibitem{Saramaki2015}
Saramäki J, Moro E. 2015 From seconds to months: an overview of multi-scale
  dynamics of mobile telephone calls.
\newblock \emph{The European Physical Journal B} \textbf{88}, 6.
\newblock (\doi{10.1140/epjb/e2015-60106-6}).

\bibitem{Ahnert2015}
Ahnert R, Ahnert SE. 2015 Protestant letter networks in the reign of mary i: A
  quantitative approach.
\newblock \emph{{ELH}} \textbf{82}, 1, 1--1.
\newblock (\doi{10.1353/elh.2015.0000}).

\bibitem{Holme2019}
Holme P. 2019 Rare and everywhere: Perspectives on scale-free networks.
\newblock \emph{Nature Communications} \textbf{10}, 1.
\newblock (\doi{10.1038/s41467-019-09038-8}).

\bibitem{Granovetter1973}
Granovetter MS. 1973 The strength of weak ties.
\newblock \emph{American Journal of Sociology} \textbf{78}, 6, 1360--1380.
\newblock (\doi{10.1086/225469}).

\bibitem{Carmines1982}
Zeller RA, Nock SL, Carmines EG. 1982 Measurement in the social sciences: The
  link between theory and data.
\newblock \emph{Contemporary Sociology} \textbf{11}, 1, 79.
\newblock (\doi{10.2307/2066656}).

\bibitem{Urena2020}
Ure{\~{n}}a-Carrion J, Saram\"{a}ki J, Kivel\"{a} M. 2020 Estimating tie
  strength in social networks using temporal communication data.
\newblock \emph{{EPJ} Data Science} \textbf{9}, 1.
\newblock (\doi{10.1140/epjds/s13688-020-00256-5}).

\bibitem{Navarro2017}
Navarro H, Miritello G, Canales A, Moro E. 2017 Temporal patterns behind the
  strength of persistent ties.
\newblock \emph{{EPJ} Data Science} \textbf{6}, 1.
\newblock (\doi{10.1140/epjds/s13688-017-0127-3}).

\bibitem{Goh2008}
Goh KI, Barab{\'{a}}si AL. 2008 Burstiness and memory in complex systems.
\newblock \emph{{EPL} (Europhysics Letters)} \textbf{81}, 4, 48002.
\newblock (\doi{10.1209/0295-5075/81/48002}).

\bibitem{KarsaiCorrelated}
Karsai M, Kaski K, Kert{\'{e}}sz J. 2012 Correlated dynamics in egocentric
  communication networks.
\newblock \emph{{PLoS} {ONE}} \textbf{7}, 7, e40612.
\newblock (\doi{10.1371/journal.pone.0040612}).

\bibitem{malmgren2008poissonian}
Malmgren RD, Stouffer DB, Motter AE, Amaral LA. 2008 A poissonian explanation
  for heavy tails in e-mail communication.
\newblock \emph{Proceedings of the National Academy of Sciences} \textbf{105},
  47, 18153--18158.

\bibitem{Oliveira2005}
Oliveira JG, Barab{\'{a}}si AL. 2005 Darwin and einstein correspondence
  patterns.
\newblock \emph{Nature} \textbf{437}, 7063, 1251--1251.
\newblock (\doi{10.1038/4371251a}).

\bibitem{Malmgren2009}
Malmgren RD, Stouffer DB, Campanharo ASLO, Amaral LAN. 2009 On universality in
  human correspondence activity.
\newblock \emph{Science} \textbf{325}, 5948, 1696--1700.
\newblock (\doi{10.1126/science.1174562}).

\bibitem{Colavizza2014}
Colavizza G. 2014 \emph{Mapping early modern news networks: a digital
  humanities approach}.
\newblock Master's thesis, Università Ca' Foscari Venezia.
\newblock (\doi{10.13140/RG.2.2.26344.88326}).
\newblock Http://hdl.handle.net/10579/4893.

\bibitem{Dooley2010}
Dooley B (ed.). 2010 \emph{The Dissemination of News and the Emergence of
  Contemporaneity in Early Modern Europe}.
\newblock Ashgate Farham.
\newblock (\doi{10.4324/9781315240244}).

\bibitem{Viswanath2009}
Viswanath B, Mislove A, Cha M, Gummadi KP. 2009 {On the evolution of user
  interaction in Facebook}.
\newblock In: \emph{Proceedings of the 2nd {ACM} workshop on Online social
  networks - WOSN 09}. {ACM} Press.
\newblock (\doi{10.1145/1592665.1592675}).

\bibitem{nraaai15}
Rossi RA, Ahmed NK. 2015 The network data repository with interactive graph
  analytics and visualization.
\newblock In: \emph{AAAI}.

\bibitem{Karimi2014}
Karimi F, Ramenzoni VC, Holme P. 2014 Structural differences between open and
  direct communication in an online community.
\newblock \emph{Physica A: Statistical Mechanics and its Applications}
  \textbf{414}, 263--273.
\newblock (\doi{10.1016/j.physa.2014.07.037}).

\bibitem{dumont2016}
Dumont S. 2016 {correspSearch --– Connecting Scholarly Editions of Letters}.
\newblock \emph{Journal of the Text Encoding Initiative} \textbf{6}, 10.
\newblock (\doi{10.4000/jtei.1742}).

\bibitem{hyvonen-et-al-ldf-2014}
Hyv\"{o}nen E, Tuominen J, Alonen M, M\"{a}kel\"{a} E. 2014 {Linked Data
  Finland}: {A} 7-star model and platform for publishing and re-using linked
  datasets.
\newblock In: \emph{The Semantic Web: {ESWC} 2014 Satellite Events, Revised
  Selected Papers}, pp. 226--230. Springer-Verlag.

\bibitem{heath-bizer-2011}
Heath T, Bizer C. 2011 \emph{Linked Data: Evolving the Web into a Global Data
  Space}.
\newblock Synthesis Lectures on the Semantic Web: Theory and Technology. Morgan
  \& Claypool.

\bibitem{Kivela2015}
Kivel\"{a} M, Porter MA. 2015 Estimating interevent time distributions from
  finite observation periods in communication networks.
\newblock \emph{Physical Review E} \textbf{92}, 5.
\newblock (\doi{10.1103/physreve.92.052813}).

\bibitem{hyvonen-et-al-lettersampo-2021}
Hyv\"{o}nen E, Leskinen P, Tuominen J. 2020.
\newblock {From the Republic of Letters to LetterSampo -- Historical Letters on
  the Semantic Web}.
\newblock White paper, Aalto University, Semantic Computing Research Group
  (SeCo):
  \url{https://seco.cs.aalto.fi/publications/2020/hyvonen-et-al-lettersampo-2020.pdf}.

\bibitem{leskinen-et-al-lettersampo-2021}
Leskinen P, Ureña-Carrion J, Tuominen J, Kivelä M, Hyv\"{o}nen E.
\newblock Using a linked data service for analysing temporal communication
  networks -- applications to emails and historical letters.
\newblock White paper, Aalto University, Semantic Computing Research Group
  (SeCo).

\bibitem{hyvonen-sampos-dhn-2020}
Hyvönen E. 2020 {"Sampo"} model and semantic portals for digital humanities on
  the semantic web.
\newblock In: \emph{DHN 2020 Digital Humanities in the Nordic Countries.
  Proceedings of the Digital Humanities in the Nordic Countries 5th
  Conference}, pp. 373--378. CEUR Workshop Proceedings, vol. 2612.

\end{thebibliography}

\end{document}